\documentclass[journal]{IEEEtran}
\usepackage{fancyhdr}

\usepackage[acronyms,nonumberlist,nopostdot,nomain,nogroupskip]{glossaries}
\usepackage{amsmath,amsfonts}
\usepackage{algorithmic}
\usepackage{algorithm}
\usepackage{array}
\usepackage{pifont}
\usepackage{multirow}
\newcommand{\cmark}{\textcolor{green!50!black}{\ding{51}}}
\newcommand{\xmark}{\textcolor{red!50!black}{\ding{55}}}
\usepackage[caption=false,font=normalsize,labelfont=sf,textfont=sf]{subfig}
\usepackage{textcomp}
\usepackage{stfloats}
\usepackage{url}
\usepackage{verbatim}
\usepackage{graphicx}
\usepackage{cite}
\usepackage{tabularx}
\usepackage{makecell}
\usepackage{array}
\usepackage{colortbl}
\usepackage{siunitx}
\usepackage{tikz}
\usepackage{booktabs}
\usepackage{graphicx}
\usepackage{textcomp}
\usepackage{xcolor}
\usepackage{xcolor}

\newcommand{\beq}{\begin{equation}}
\newcommand{\eeq}{\end{equation}}
\newcommand{\bit}{\begin{itemize}}
\newcommand{\eit}{\end{itemize}}

\newcommand\diagfil[4]{%
  \multicolumn{1}{p{#1}}{\hskip-\tabcolsep
  $\vcenter{\begin{tikzpicture}[baseline=0,anchor=south west,inner sep=0pt,outer sep=0pt]
  \path[use as bounding box] (0,0) rectangle (#1+2\tabcolsep,\baselineskip);
  \node[minimum width={#1+2\tabcolsep},minimum height=\baselineskip+\extrarowheight+0+0,fill=#2] (box)at(0,0) {};
  \fill [#3] (box.south west)--(box.north east)|- cycle;
  \node[anchor=center] at (box.center) {#4};
  \end{tikzpicture}}$\hskip-\tabcolsep}}

\begin{document}

\title{A CSI Dataset for Wireless Human Sensing\\on 80~MHz Wi-Fi Channels}

\author{\IEEEauthorblockN{Francesca Meneghello$^*$, Nicol\`o Dal Fabbro$^*$, Domenico Garlisi$^{\dag\S}$, Ilenia Tinnirello$^{\ddag\S}$, Michele Rossi$^{*\#}$}\\
 \IEEEauthorblockA{$^*$ Department of Information Engineering, University of Padova, Italy}\\
\IEEEauthorblockA{$^{\dag}$ Department of Mathematics and Computer Science, University of Palermo, Italy}\\
 \IEEEauthorblockA{$^{\ddag}$ Department of Engineering, University of Palermo, Italy}\\
 \IEEEauthorblockA{$^{\S}$ CNIT, Padova, Italy}\\
 \IEEEauthorblockA{$\#$ Department of Mathematics ``Tullio Levi-Civita'', University of Padova, Italy}\\
 \vspace{-0.6cm}
 }
\thispagestyle{fancy}

\maketitle

\begin{abstract}
In the last years, several machine learning-based techniques have been proposed to monitor human movements from Wi-Fi channel readings. However, the development of {\it domain-adaptive} algorithms that robustly work across different environments is still an open problem, whose solution requires large datasets characterized by {\it strong domain diversity}, in terms of environments, persons and \mbox{Wi-Fi} hardware. To date, the few public datasets available are mostly obsolete -- as obtained via \mbox{Wi-Fi} devices operating on 20 or 40~MHz bands -- and contain little or no domain diversity, thus dramatically limiting the advancements in the design of sensing algorithms. The present contribution aims to fill this gap by providing a dataset of IEEE~802.11ac channel measurements over an 80~MHz bandwidth channel featuring notable domain diversity, through measurement campaigns that involved thirteen subjects across different environments, days, and with different hardware. Novel experimental data is provided by blocking the direct path between the transmitter and the monitor, and collecting measurements in a semi-anechoic chamber (no multi-path fading). Overall, the dataset -- available on IEEE DataPort~\cite{dataset_CSI} -- contains more than thirteen hours of channel state information readings (23.6~GB), allowing researchers to test activity/identity recognition and people counting algorithms.\vspace{-0.3cm} 
\end{abstract}
\begin{IEEEkeywords}
\mbox{Wi-Fi} sensing, channel frequency response.\vspace{-0.3cm}
\end{IEEEkeywords}


\section{Introduction}
\thispagestyle{fancy}

In recent years, spurred by the pervasiveness of \mbox{Wi-Fi} devices, wireless human sensing strategies that rely on movement-induced modifications to the propagation of radio signals have been widely investigated~\cite{Chen2023}. Being a passive -- and thus non-intrusive -- approach, \mbox{Wi-Fi-based} human sensing (WHS) has attracted considerable attention for different applications such as, e.g., device-free monitoring~\cite{GaitSenseDS} and activity recognition (AR)~\cite{SHARP-PAPER}. Moreover, by leveraging existing communication infrastructures, WHS grants an easy and low-cost deployment.

In this work, we present a large dataset for WHS, which aims at providing researchers with a means to develop new algorithms and assess their performance on common data, thus allowing for fair comparisons. The availability of large-scale and comprehensive datasets is especially necessary when considering data-driven algorithms exploiting machine/deep learning techniques. These methods are adopted by the vast majority of experimental WHS research, as model-based solutions are of difficult implementation in a wireless context, given the complexity of radio signal propagation.
However, public WHS datasets are still very few and often outdated: many of the measurement campaigns that were made available in recent years were performed on $20$ or $40$~MHz frequency bands, and do not contain sufficient {\it domain diversity}.

To bridge this gap, the dataset we made publicly available on IEEE DataPort~\cite{dataset_CSI} has been collected through IEEE~802.11ac devices operating on $80$~MHz bands. Moreover, in contrast with currently available datasets, it contains data featuring notable domain diversity across measurement days, environments, people and hardware. This will allow researchers to assess the performance of their algorithms in completely different situations than those considered at training time, and, in turn, to develop strategies to address the domain adaptation problem in WHS. In fact, although some preliminary WHS approaches were successful, we advocate that some fundamental challenges are still open and, among these, one of the most relevant is the design of robust algorithms, i.e., able to generalize and adapt to \emph{new domains}. The dataset contains data for developing AR, person identification (PI) and people counting (PC) applications, providing Wi-Fi channel measurements for up to seven activities, ten people, and seven environments. Data collected in a semi-anechoic chamber, emulating an open-area-test site, are also included in the dataset. 
Moreover, we provide for the first time data collected when the direct path between transmitter and monitor is completely obstructed.
Part of the data has been used to validate SHARP, our recent algorithm for environment- and person-independent \mbox{Wi-Fi-based} AR~\cite{SHARP-PAPER}.

\section{Background and Existing Datasets}
\label{sec:background_datasets}

In September 2020, the IEEE~802.11bf working group was established to empower Wi-Fi devices with {\it sensing capabilities}~\cite{Chen2023}. The goal is to allow Wi-Fi routers to perform the dual role of {\it communication access points} (AP) and {\it monitoring devices}, leveraging ongoing Wi-Fi traffic as well as ad-hoc packets to deliver sensing services. Note that the new IEEE~802.11bf standard is not expected to define specific use cases or algorithms for sensing that, in turn, remain open to implementation and require further investigation by the research community~\cite{meneghello2023toward}. Wi-Fi-based sensing exploits the fact that the presence and the movement of objects in the propagation environment affect the Wi-Fi signal (multi-path) propagation, and these modifications can be estimated via dedicated signal processing on the Wi-Fi channel frequency response (CFR) -- also referred to as channel state information (CSI).
\mbox{Wi-Fi} systems adopt orthogonal frequency division multiplexing (OFDM) by transmitting over partially overlapping and orthogonal \mbox{sub-channels}, and the CFR is continuously estimated for all of them.
Thus, for each pair of transmitting and receiving antennas, the CFR consists of a vector of complex numbers specifying the attenuation and phase shift experienced by the signal over each OFDM \mbox{sub-channel}. Interested readers can find a complete description of the \mbox{Wi-Fi} channel in~\cite{SHARP-PAPER}.

\subsection{Comparison against other public datasets}

\begin{table*}[t]
    \centering
    \resizebox{2\columnwidth}{!}{
    \begin{tabular}{cccccccccc}  
    \toprule
    \multirow{2}{*}{\textbf{dataset}} & \textbf{no. days per} & \multirow{2}{*}{\textbf{no. environments}} & \textbf{no. involved} & \textbf{no. concurrent} & \multirow{2}{*}{\textbf{no. activities}} & \textbf{no. different } & \textbf{obstructed} & \multirow{2}{*}{\textbf{standard}} & \multirow{2}{*}{\textbf{bandwidth}}  \\
     & \textbf{environment} &  & \textbf{people} & \textbf{people} &  & \textbf{Tx hardware} & \textbf{direct path} &   \\
    \midrule
    AR\cite{FirstDS} & 1 & 1 & 6 & 1 & 6 & 1 & \xmark & 802.11n &20MHz \ \\
    \midrule
    AR\cite{Wiar} & 1 & 3 & 10 & 1 & 16 & 1 & \xmark & 802.11n &20MHz \\
    \midrule
    AR\cite{LearningCSI_tool} & 1 & 3 & 1 & 1 & 7 & 1 & \xmark & 802.11n &40MHz \\
    \midrule
    AR\cite{RasbPiApart} & 1 & 4 & 1 & 1 & 11 & 1 & \xmark & 802.11ac &80MHz \\
    \midrule
    AR\cite{Restuccia} & 1 & 3 & 1 & 1 & 4 & 1 & \xmark & 802.11ac & 80MHz \\
    \midrule
    AR\cite{DifferentHardw}& 1 & 3 & 1 & 1 & 5-8 & 2 & \xmark & 802.11ac &40-80MHz \\
    \midrule
    PC\cite{diDomenico2016}& 1 & 3 & 7 & 1-7 & 1 & 1 & \xmark & 802.11n &20MHz \\
    \midrule
    PI\cite{GaitSenseDS}& 1 & 2 & 11 & 1 & 1 & 1 & \xmark  & 802.11n &20MHz \\
    \midrule
    \cellcolor{blue!10}\textbf{AR-PC-PI \cite{dataset_CSI}} & \cellcolor{blue!10} & \cellcolor{blue!10}\textbf{7 (including a} & \cellcolor{blue!10} & \cellcolor{blue!10} & \cellcolor{blue!10} & \cellcolor{blue!10} & \cellcolor{blue!10} & \cellcolor{blue!10} & \cellcolor{blue!10} \\
    \cellcolor{blue!10}\textbf{(our)} & \multirow{-2}{*}{\cellcolor{blue!10}\textbf{1-5}} & \cellcolor{blue!10}\textbf{semi-anechoic chamber)} & \multirow{-2}{*}{\cellcolor{blue!10}\textbf{13}} & \multirow{-2}{*}{\cellcolor{blue!10}\textbf{1-10}} & \multirow{-2}{*}{\cellcolor{blue!10}\textbf{1-7}} & \multirow{-2}{*}{\cellcolor{blue!10}\textbf{3}} & \multirow{-2}{*}{\cellcolor{blue!10}\textbf{\cmark}} & \multirow{-2}{*}{\cellcolor{blue!10}\textbf{802.11ac}}  & \multirow{-2}{*}{\cellcolor{blue!10}\textbf{80MHz}}\\
    \bottomrule
\end{tabular}} \vspace{0.2cm}
    \caption{Public datasets for Wi-Fi-based AR, PC and PI.\vspace{-0.5cm}}
    \label{tab:Comparison}
\end{table*}

Although all commercial \mbox{Wi-Fi} chipsets estimate the CFR, manufacturers do not make this data easily accessible for any other use different from communication. To overcome this, over the years, researchers have designed and implemented three tools to extract the (estimated) CFR. Two of them, namely Linux CSI~\cite{Halperin2011} and Atheros CSI~\cite{Xie2015}, target network interface cards (NICs) implementing IEEE~802.11n (operating over $20$ or $40$~MHz frequency bands). The third one, Nexmon CSI~\cite{Gringoli2019}, allows extracting the CFR from specific \mbox{Wi-Fi} chipsets implementing IEEE~802.11ac.
Leveraging such tools, several experimental campaigns have been realized in recent years, and have been used by researchers to develop different WHS algorithms. However, releasing the collected data has not been a common practice so far and, in turn, public datasets for WHS are still very few.

In~\cite{FirstDS,Wiar}, two datasets populated with IEEE~802.11n CFR data captured over thirty OFDM sub-channels have been released. Six people carrying out six activities across a single environment, and ten volunteers, sixteen activities, and three environments are respectively considered in~\cite{FirstDS} and~\cite{Wiar}. 
The work in~\cite{LearningCSI_tool} provides a dataset collected over 114 OFDM sub-channels as a single person performs six activities in three environments. In~\cite{RasbPiApart}, CFR data was collected using Raspberry-Pi with Nexmon, for 242 OFDM sub-channels (on $80$~MHz bands), while a subject performed activities in four rooms.
Asus RT-AC86U routers empowered with Nexmon CSI are used in~\cite{Restuccia}, where a single subject performs activities across three environments.
In~\cite{DifferentHardw}, a dataset featuring a single subject moving in three environments has been collected using a Raspberry-Pi and an Asus \mbox{RT-AC86U} router. 
As for data enabling people counting applications, the dataset presented in~\cite{diDomenico2016} provides measurements of up to seven people. However, it only contains measurement via devices working with the IEEE~802.11n standard over a $20$~MHz channel. Finally, to the best of our knowledge, the only publicly available dataset for human identification through \mbox{Wi-Fi} was released as part of the gait recognition study in~\cite{GaitSenseDS}. CFR data was collected from eleven volunteers in two environments over a $20$~MHz channel.

In this article, we present a comprehensive dataset providing experimental IEEE~802.11ac CFR data collected across several different environments (space diversity), persons, and transmission hardware (hardware diversity). Our dataset is the first that includes data collected in the same environment on different days (time diversity). This is of high value, as it enables the assessment of the impact that small environmental changes -- which are likely to happen in real-world scenarios -- have on sensing applications (as detailed in Section~\ref{example-application}). The dataset has been collected over an $80$~MHz frequency band to allow designing algorithms that can be integrated into modern Wi-Fi devices, as those currently deployed in homes and buildings. Moreover, since the sensing resolution increases with the bandwidth~\cite{FirstDS}, our dataset allows developing more advanced sensing algorithms with respect to previous datasets available in the literature. We also remark that, by selecting the CFR values of sub-sets of the collected $256$ OFDM sub-channels at $80$~MHz, as specified in the IEEE~802.11ac standard, one can still design sensing algorithms working on the $40$~MHz and $20$~MHz sub-bands used by older Wi-Fi systems.
To our knowledge, this is the first dataset providing data for PC and PI from an $80$~MHz bandwidth communication link, and the first to provide a high domain diversity for AR.
For the first time, we also provide data for measurement campaigns performed when the direct path between the transmitter and the monitor is completely obstructed, and AR measurements in a semi-anechoic chamber where the multi-path fading is strongly reduced, to enable the evaluation of the impact of obstacles in the sensing environment and interference on the performance of WHS algorithms.
A detailed comparison among our dataset and existing AR, PI, and PC datasets is provided in Tab~\ref{tab:Comparison}.


\section{Experimental Setup}
\label{esperimental-setup}

In the following, we summarize the methodology used to collect data. The complete workflow is in \cite{dataset_CSI} for replicability.

\subsection{Wi-Fi network setup and CFR data collection}\label{subsec:network_setup}

We considered a \mbox{Wi-Fi} network consisting of a single communication link between two routers, one acting as the AP and the other as a station connected to it. This choice does not imply any loss of generality, as wireless sensing can be performed by the monitor node considering a single channel at a time. In the presence of multiple radio links, the monitor station may scan the available links according to some time-division strategy. We used Wi-Fi routers as they allow for better control of the experimental setup with respect to implementing the \texttt{Tx} and the \texttt{Rx} with other Wi-Fi-enabled devices (e.g., computers or smartphones). 
The network was set up using \texttt{OpenWrt} to operate on the IEEE~802.11ac channel number $42$, with a central frequency of $5,210$~MHz and $80$~MHz of bandwidth.

Wi-Fi traffic was generated through the \texttt{iPerf3} tool setting the packet transmitting rate to $173$ packets per second, obtaining a new channel estimate every $T_c \simeq 6 \times 10^{-3}$~s, which represents a good channel sampling interval for sensing. We used modulation and coding scheme (MCS)~$4$ without frame aggregation. A single antenna was purposely enabled on the \texttt{Tx} and the \texttt{Rx} to enforce the communication over a {\it single spatial stream}. This conforms to the measurement procedure adopted in all CSI datasets for WHS and leads to better CFR estimates at the monitor, as there is no cross-interference among transmitting and receiving antennas, which would be present in a multi-antenna configuration. It follows that the resulting dataset can be used to train learning algorithms for sensing without having to apply any interference cancellation. 

The monitor (\texttt{M}) continuously estimated the CFR of the \texttt{Tx}-\texttt{M} link by capturing the packets transmitted over-the-air by the \texttt{Tx}. In all the environments but the semi-anechoic chamber, other Wi-Fi networks coexisted with our setup, causing radio interference beyond our control. As IEEE~802.11ac networks leverage the carrier sense multiple access with collision avoidance (CSMA/CA), once interference occurs, no sensing packets are collected. The lack or delay in the acquisition of CFR samples mirrors a real-world scenario and allows evaluating the impact of the phenomenon.

\subsection{Hardware specifications}\label{subsec:hardware}

Three different setups for the \texttt{Tx}-\texttt{Rx} routers were considered. Two Netgear X4S AC2600 IEEE 802.11ac routers were used as \texttt{Tx}-\texttt{Rx} in the first setup, whereas two Asus RT-AC86U IEEE 802.11ac routers were deployed for the second. The third setup consisted of a Netgear X4S AC2600 and a TP-Link AD7200 IEEE 802.11ac/ad router.
An Asus RT-AC86U router equipped with $N_{\rm ant}=4$ antennas was used as monitor device in all the setups. The dataset was collected using the Nexmon-CSI extraction tool~\cite{Gringoli2019}.

\subsection{Measurement setup}\label{subsec:workflow}

The dataset has been collected by deploying the experimental \mbox{Wi-Fi} network (\texttt{Tx}-\texttt{Rx} pair and monitor) in seven different environments, i.e., a bedroom, a living room, a kitchen, a university laboratory, a university office, a semi-anechoic chamber, and a meeting room, by also changing the respective positions of \texttt{Tx}, \texttt{Rx} and monitor devices. 
We considered two types of direct-path obstructions to enable the evaluation of sensing applications in non-line-of-sight scenarios: a wood bookcase and a concrete block wall (see Section~\ref{sec:dataset-organiz}).

\begin{figure}[t!]
\centering
		\includegraphics[width=0.9\columnwidth]{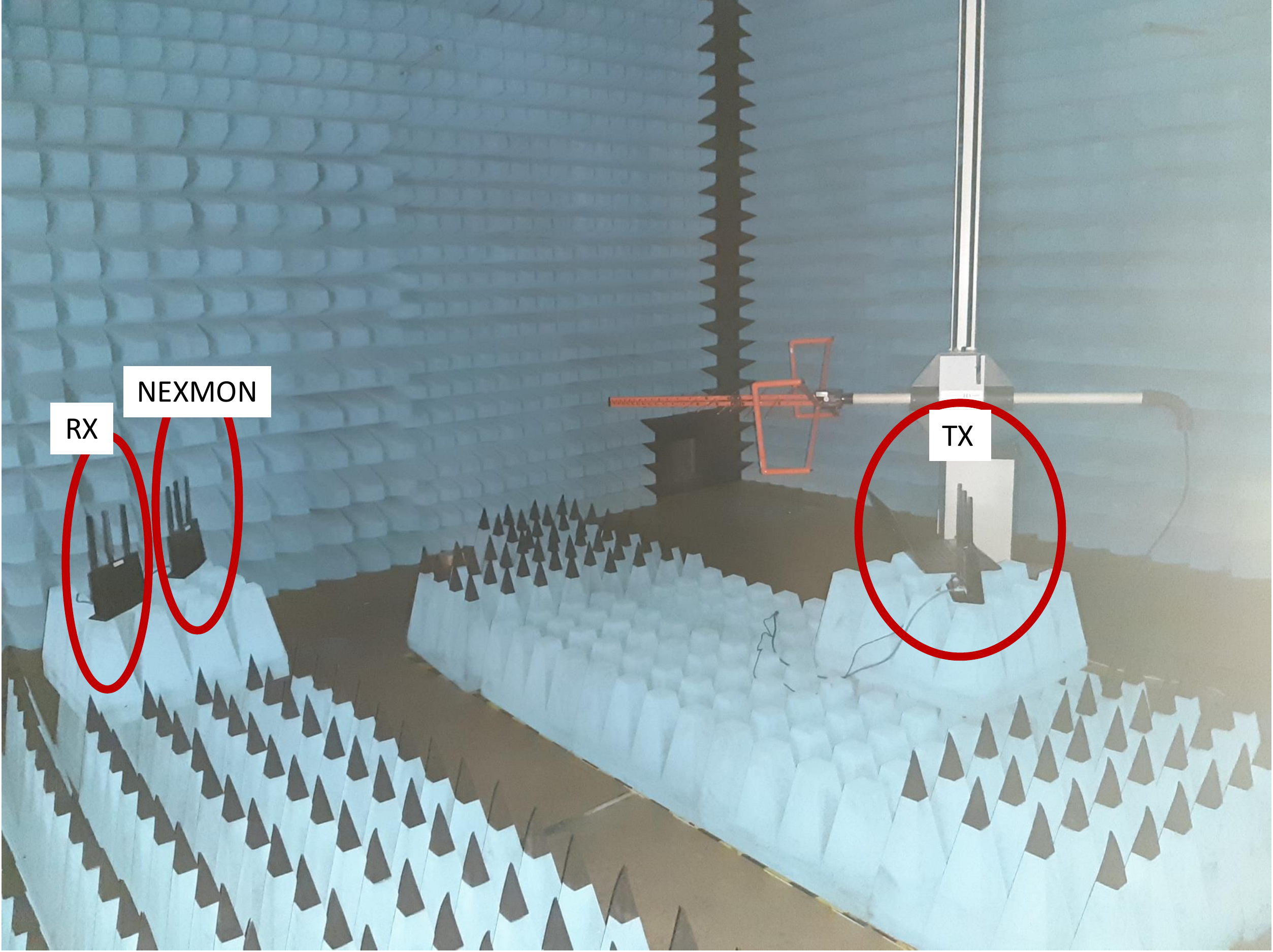}
 		\setlength\abovecaptionskip{-.05cm}
		\caption{Semi-anechoic chamber. The \texttt{Tx}, \texttt{Rx} and monitor Asus routers used for data transmission and CFR collection are indicated in the picture.\vspace{-0.4cm}}
		\label{fig:semianecoic-picture}
\end{figure}

As an example of the data collection campaign, Fig.~\ref{fig:semianecoic-picture} shows the experimental setup deployed in the semi-anechoic chamber and consisting of three Asus routers. The walls of the semi-anechoic chamber are fully covered by $40$-$50$~cm long pyramidal radio-absorbing panels which guarantee an absorbing factor of $110$~dB at the considered frequencies ($1$-$10$~GHz). Movable radio-absorbing panels partially cover the floor while keeping enough uncovered space for users' movement. The semi-anechoic chamber did not contain any objects and no reflectors were used during the experiments.

\section{Data Description}\label{sec:data-description}

Our dataset contains more than thirteen hours of CFR collections, resulting in $23.6$~GB of data.
In the following, the sequence of CFR vectors associated with an acquisition (one \texttt{.mat} file in the dataset) is referred to as {\it CFR trace}.
Each CFR trace is saved as a $(N * N_{\rm ant}) \times M$ dimensional complex matrix, where each row is a CFR vector and $M$ is the number of monitored OFDM sub-channels. 
Each trace contains data collected during $40 - 300$ seconds, resulting in around $N = 6,600 - 50,000$ CFR vectors per monitor antenna per trace. As our dataset is collected considering transmissions on an $80$~MHz band, $256$ sub-channels can, in principle, be monitored. However, Nexmon only returns the CFR for the $M=242$ data sub-channels while no information is provided for the control sub-channels. As part of our previous work, we made available a Python script to further process the data~\cite{SHARP-PAPER}.

\noindent \textbf{Dataset domain diversity.} The campaigns have been performed on different days through several months (April-December, $2020$ and January-September, $2022$), ensuring considerable time diversity. Thirteen volunteers were involved: ten males, in the following, referred to as \texttt{P1}, \texttt{P4}, \texttt{P6}-\texttt{P13} and three females, indicated with \texttt{P2}, \texttt{P3} and \texttt{P5}. Four volunteers performing activities within six different environments were considered for AR and two different Wi-Fi network configurations were deployed to include hardware diversity in the data. For what concerns the PI and PC applications, we provide diversity in terms of people involved (ten), and \mbox{Wi-Fi} hardware (four different setups). For all three applications, domain diversity in terms of LOS/NLOS conditions for the transmitter-monitor link is included.

\begin{table*}[t!]
\centering
\resizebox{2\columnwidth}{!}{
\begin{tabular}{cccccccccc}
\toprule
\textbf{set} & \textbf{campaigns} & \textbf{environment} & \textbf{w $\times$ l $\times$ h [m]}& \textbf{obst.} & \textbf{devices pos.} & \textbf{\texttt{Tx}} & \textbf{\texttt{Rx}} & \textbf{person, \texttt{Pi}} & \textbf{furniture}  \\ \midrule
\textbf{\texttt{AR-1}} & \textbf{\texttt{a}}-\textbf{\texttt{b}}-\textbf{\texttt{c}}-\textbf{\texttt{d}}-\textbf{\texttt{e}} & bedroom & 5 $\times$ 6 $\times$ 4 & - & \cellcolor{green!15}\texttt{M1}-\texttt{Tx}-\texttt{Rx} & Netgear & Netgear & \cellcolor{gray!10}\texttt{P1} & \multirow{4}{*}{bookcase, 2 beds, desk, chairs} \\
\textbf{\texttt{AR-2}} & \textbf{\texttt{a}} & bedroom& 5 $\times$ 6 $\times$ 4 & - & \cellcolor{green!15}\texttt{M1}-\texttt{Tx}-\texttt{Rx} & Netgear & Netgear & \cellcolor{gray!25}\texttt{P2} & \\ 
\textbf{\texttt{AR-3}} & \textbf{\texttt{a}}-\textbf{\texttt{b}} & bedroom & 5 $\times$ 6 $\times$ 4 & \cmark &  \diagfil{1.6cm}{red!20}{green!15}{\texttt{M2}-\texttt{Tx}-\texttt{Rx}} & Netgear & Netgear & \cellcolor{gray!10}\texttt{P1}& \\
\textbf{\texttt{AR-4}} & \textbf{\texttt{a}} & bedroom & 5 $\times$ 6 $\times$ 4 & \cmark & \diagfil{1.6cm}{red!20}{green!15}{\texttt{M2}-\texttt{Tx}-\texttt{Rx}} & Netgear & Netgear & \cellcolor{gray!25}\texttt{P2}& \\
\textbf{\texttt{AR-5}} & \textbf{\texttt{a}}-\textbf{\texttt{b}} & living room & 5 $\times$ 6 $\times$ 4 & - & \cellcolor{yellow!50}\texttt{M3}-\texttt{Tx1}-\texttt{Rx1} & Netgear & Netgear & \cellcolor{gray!10}\texttt{P1} & armchair, TV cabinet, table, chairs, 2 sofas\\
\textbf{\texttt{AR-6}} & \textbf{\texttt{a}} & kitchen & 3.5 $\times$ 3 $\times$ 3.2 & - & \cellcolor{yellow!50}\texttt{M3}-\texttt{Tx1}-\texttt{Rx1} & Netgear & Netgear & \cellcolor{gray!10}\texttt{P1} & hob, 2 cabinets\\
\textbf{\texttt{AR-7}} & \textbf{\texttt{a}} & laboratory & 7.5 $\times$ 3.5 $\times$ 2.9 & - & \cellcolor{yellow!50}\texttt{M3}-\texttt{Tx1}-\texttt{Rx1} & Netgear & Netgear & \cellcolor{gray!40}\texttt{P3} & 4 desks, 6 workstations, 6 displays\\
\textbf{\texttt{AR-8}} & \textbf{\texttt{a}}-\textbf{\texttt{b}} & office&  4 $\times$ 6 $\times$ 3 & - & \cellcolor{yellow!50}\texttt{M3}-\texttt{Tx1}-\texttt{Rx1} & Asus & Asus & \cellcolor{gray!60}\texttt{P4} & desk, workstation, display \\
\textbf{\texttt{AR-9}} & \textbf{\texttt{a}}-\textbf{\texttt{b}}-\textbf{\texttt{c}} & semi-anechoic & 9 $\times$ 7 $\times$ 3.4 & - & \cellcolor{yellow!50}\texttt{M3}-\texttt{Tx1}-\texttt{Rx1} & Asus & Asus & \cellcolor{gray!60}\texttt{P4} & no furniture\\
\midrule
\textbf{\texttt{PI-1}} & \textbf{\texttt{a}} & meeting room & 7 $\times$ 7.5 $\times$ 3.5 & - & \cellcolor{yellow!50}\texttt{M3}-\texttt{Tx1}-\texttt{Rx1} & Netgear & Netgear & \cellcolor{gray!80}\texttt{P3,P5-P13} & \multirow{4}{*}{7 desks, chairs}\\
\textbf{\texttt{PI-2}} & \textbf{\texttt{a}} & meeting room & 7 $\times$ 7.5 $\times$ 3.5 & \cmark & \cellcolor{yellow!50}\texttt{M3}-\texttt{Tx2}-\texttt{Rx2} & Netgear & TP-Link & \cellcolor{gray!80}\texttt{P3,P5-P13} & \\
\textbf{\texttt{PI-3}} & \textbf{\texttt{a}} & meeting room & 7 $\times$ 7.5 $\times$ 3.5 & - & \cellcolor{yellow!50}\texttt{M4}-\texttt{Tx1}-\texttt{Rx1} & Netgear & Netgear & \cellcolor{gray!80}\texttt{P3,P5-P13} & \\
\textbf{\texttt{PI-4}} & \textbf{\texttt{a}} & meeting room & 7 $\times$ 7.5 $\times$ 3.5 & \cmark & \cellcolor{yellow!50}\texttt{M4}-\texttt{Tx2}-\texttt{Rx2} & Netgear & TP-Link & \cellcolor{gray!80}\texttt{P3,P5-P13} & \\
\midrule
\textbf{\texttt{PC-1}} & \textbf{\texttt{a}} & meeting room & 7 $\times$ 7.5 $\times$ 3.5 & - & \cellcolor{yellow!50}\texttt{M3}-\texttt{Tx1}-\texttt{Rx1} & Netgear & Netgear & \cellcolor{gray!80}\texttt{P3,P5-P13} & \multirow{4}{*}{7 desks, chairs}\\
\textbf{\texttt{PC-2}} & \textbf{\texttt{a}} & meeting room & 7 $\times$ 7.5 $\times$ 3.5 & \cmark & \cellcolor{yellow!50}\texttt{M3}-\texttt{Tx2}-\texttt{Rx2} & Netgear & TP-Link & \cellcolor{gray!80}\texttt{P3,P5-P13} & \\
\textbf{\texttt{PC-3}} & \textbf{\texttt{a}} & meeting room & 7 $\times$ 7.5 $\times$ 3.5 & - & \cellcolor{yellow!50}\texttt{M4}-\texttt{Tx1}-\texttt{Rx1} & Netgear & Netgear & \cellcolor{gray!80}\texttt{P3,P5-P13} & \\
\textbf{\texttt{PC-4}} & \textbf{\texttt{a}} & meeting room & 7 $\times$ 7.5 $\times$ 3.5 & \cmark & \cellcolor{yellow!50}\texttt{M4}-\texttt{Tx2}-\texttt{Rx2} & Netgear & TP-Link & \cellcolor{gray!80}\texttt{P3,P5-P13} & \\
\bottomrule
\end{tabular}}\vspace{0.15cm}   
\caption{Measurement conditions. For each set, we specify the number of campaigns performed, the monitored environment and its dimensions (width $\times$ length $\times$ height), the presence of an obstacle (obst.) blocking the direct path between the transmitter and the monitor, the position of the devices (\texttt{Mj}-\texttt{Txj}-\texttt{Rxj}) and the monitored area (identified by a color as in Fig.~\ref{fig:bedAndLive}), the \texttt{Tx}-\texttt{Rx} routers brand (see Sections~\ref{subsec:hardware} for the models), the person/people (\texttt{Pi}) involved, and the environmental furniture.\label{tab:configs}\vspace{-0.7cm}}
\end{table*}

\begin{figure}[t!]
\centering
	\includegraphics[width=0.98\columnwidth]{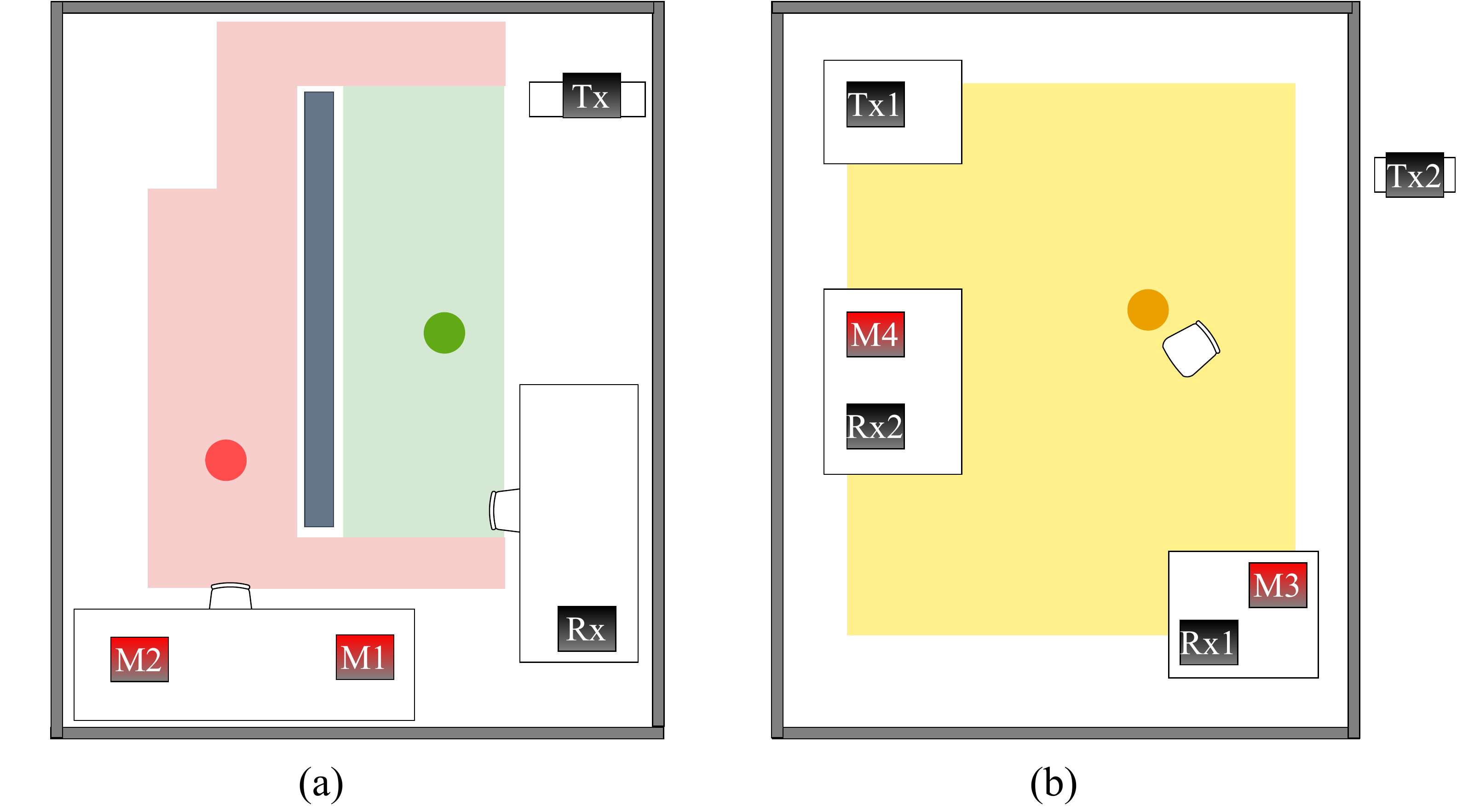}
	\setlength\abovecaptionskip{-.1cm}
	\caption{Device and user's positions in the monitored environment for (a) bedroom with a wood bookcase in the middle; (b) living room, kitchen, laboratory, office, semi-anechoic chamber, and meeting room. \texttt{Txj}, \texttt{Rxj} and \texttt{Mj}, with \texttt{j} $\in \{1, 2,3, 4\}$, denote the transmitter, the receiver and the monitor positions, respectively. The activities were performed within the colored areas as indicated in Table~\ref{tab:configs}. The chairs have been used for the still and sitting down/standing up activities. The other in-place activities were performed in the position of the dark-colored circles.
 \vspace{-0.5cm}}
		\label{fig:bedAndLive}
\end{figure}

\section{Dataset Organization}\label{sec:dataset-organiz}

The dataset is structured into 26 sub-folders, as summarized in Table~\ref{tab:configs}. Each sub-folder is identified by a prefix, a number and a letter. The prefix indicates the target application: \textbf{\texttt{AR}}, \texttt{\textbf{PI}} or \texttt{\textbf{PC}}. The number indicates a specific combination of monitored environment, hardware type and positions, measurement day and involved person/people, as described next. The letter identifies the different measurement campaigns performed for the same setting.

The first eighteen sub-folders contain CFR data intended for AR applications. The data was collected when a single person (\texttt{P1}-\texttt{P4}) is present in the monitored space. The name of each file contains a letter indicating the activity: \texttt{W} for walking, \texttt{R} for running, \texttt{J} for jumping, \texttt{L} for sitting still, \texttt{S} for standing still, \texttt{C} for sitting down/standing up and \texttt{G} for arm exercises. The CFR when no people were present in the environment is also included for each of the scenarios (label \texttt{E}).
When the activity was performed multiple times, a sequential number is present after the activity-related letter. 

The last eight sub-folders contain data for PI and PC, and were collected from ten volunteers \texttt{P3}, \texttt{P5}--\texttt{P13}. The \texttt{PI} sub-folders contain one CFR trace for each volunteer collected as they {\it moved freely} in the environment. The person identifier is included in the name of the file by a suffix starting with \texttt{p}. \texttt{\textbf{PC}} sub-folders consist of different traces, each associated with a different number of people (from $1$ to $10$) moving in the environment, as indicated in the name of each file after letter \texttt{n}. The suffixes ``\texttt{p00}'' and ``\texttt{n00}'' indicate the traces associated with the empty environment, i.e., without people. The \texttt{\textbf{PC}} sub-folders do not contain data for the single person case, as this situation is exactly the one considered for PI and, in turn, data in the \texttt{\textbf{PI}} sub-folder associated with the same measurement conditions (\texttt{Tx}-\texttt{Rx}-\texttt{M} hardware and positions) can be used.

For each set, Table~\ref{tab:configs} provides information about the monitored environment and the person/people involved ($\texttt{Pi}$), together with the position and hardware (brand) of the transmitter (\texttt{Txj}), receiver (\texttt{Rxj}) and monitor (\texttt{Mj}). 
The positions of the devices are shown in Fig.~\ref{fig:bedAndLive}-a for the bedroom and in Fig.~\ref{fig:bedAndLive}-b for the other environments. 
The colors in Table~\ref{tab:configs} and Fig.~\ref{fig:bedAndLive} indicate the areas where the activities were performed. The volunteer moved freely within the colored areas for the walking and running experiments. The sitting still and sitting down/standing up activities were performed where the chairs are positioned. The other in-place activities, i.e., jumping, standing still, and doing arm exercises, were performed in the location indicated by the dark-colored circles.

For the bedroom we considered two configurations to allow evaluating the impact of an obstruction in the environment on the sensing performance. In the former, the monitor was in position \texttt{M1} and, in turn, there existed a direct path between the transmitter and the monitor (sets \texttt{AR-1}, \texttt{AR-2}).
As for the latter, the monitor was in position \texttt{M2} so that the direct path (\texttt{Tx}-\texttt{M2}) was occluded by the bookcase in the middle of the room (gray rectangle in Fig.~\ref{fig:bedAndLive}-a), and the person was required to move in both the green and the red areas (sets \texttt{AR-3}, \texttt{AR-4}).
Two scenarios were also considered for the meeting room. The data was concurrently collected from two Wi-Fi networks, entailing \texttt{Tx1}-\texttt{Rx1} and \texttt{Tx2}-\texttt{Rx2} respectively, through two monitor devices (\texttt{M3} and \texttt{M4}). This configuration generated four simultaneous acquisitions for each \texttt{PI} (\texttt{PC}) experiment: the files with the same suffix in different sub-folders refer to simultaneous collections. The \texttt{Tx2}-\texttt{M3} and \texttt{Tx2}-\texttt{M4} paths were obstructed by the 20-cm thick concrete block wall of the meeting room (sets \texttt{PI-2}, \texttt{PI-4}, \texttt{PC-2}, \texttt{PC-4}). As the data with and without the obstruction were concurrently collected, the impact of the obstruction can be deeply evaluated.

Note that the dataset is intended for the development of AR, PI, and PC applications based on the analysis of the CFR, as shown in, e.g., \cite{SHARP-PAPER, LearningCSI_tool, Restuccia}. The precise location of the subject within the environment was not recorded making the dataset not suitable for localization and tracking purposes.

\section{Examples of Use}
\label{example-application}

\begin{figure}[t!]
\centering
		\includegraphics[width=1\columnwidth]{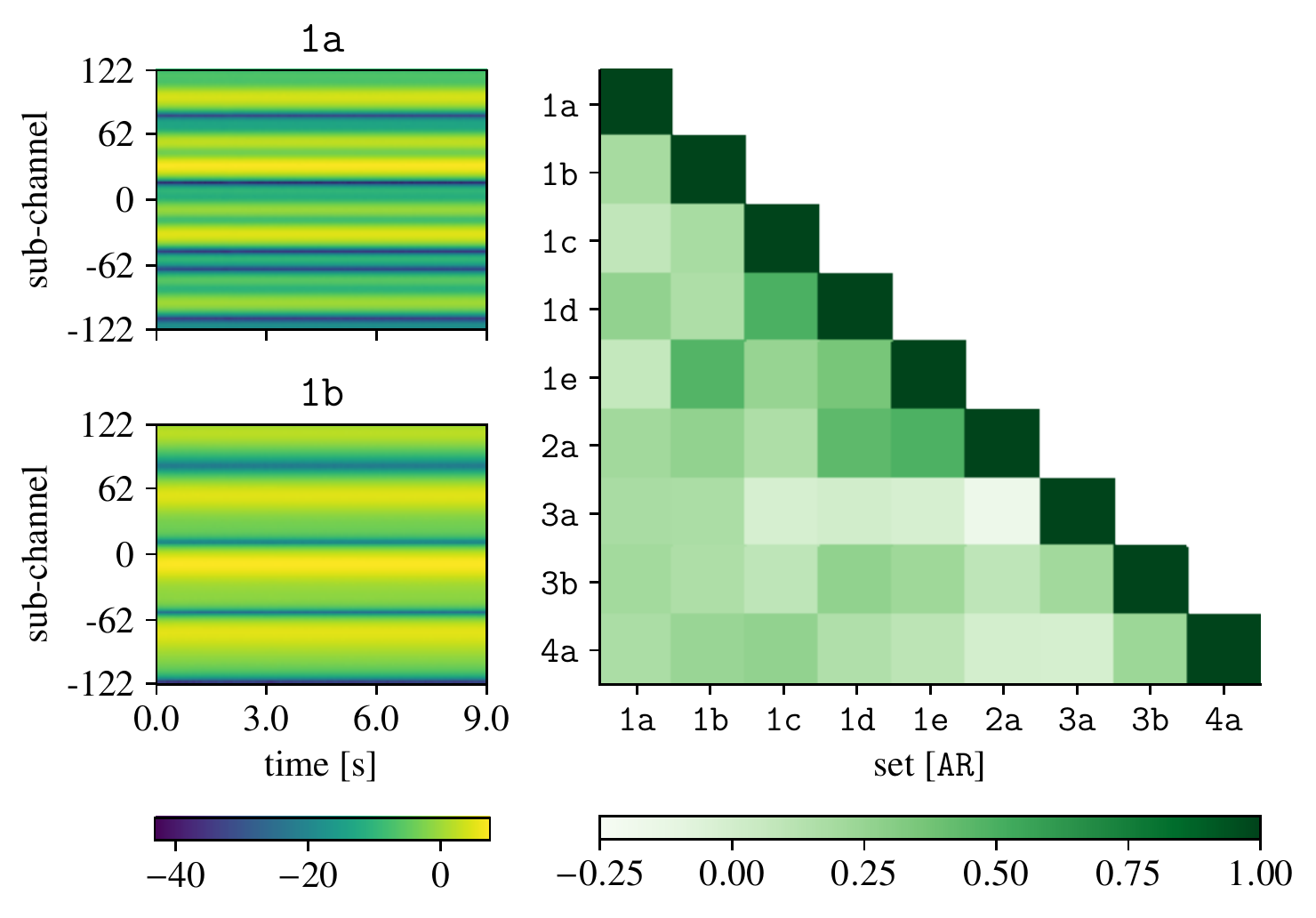}
 		\setlength\abovecaptionskip{-.5cm}
		\caption{Example of CFR amplitude collected in the empty bedroom on two different days (\texttt{AR1a\_E}, \texttt{AR1b\_E}) and Pearson coefficient computed between each pair of traces collected in the empty bedroom on different days (\texttt{AR} sets). \vspace{-0.4cm}}
		\label{fig:pearsonr}
\end{figure}

The dataset enables the effective design of several smart building applications. The AR sets can be used to design applications for assisted living and smart entertainment. The PI sets can be used to design solutions for intrusion detection and keyless access to access-restricted areas. Finally, the PC sets enable controlling the number of persons that can concurrently share the same environment to guarantee safe inter-person distance.
The dataset allows implementing algorithms that generalize over different domains such as the person, the environment, and the day of measurement. This feature is of paramount importance for the integration of sensing functionalities within commercial Wi-Fi devices, to provide plug-and-play sensing solutions. Note that the characteristics of the Wi-Fi channel strictly depend on the environment, and the channel variability is person-specific~[5]. Moreover, small changes in the displacement of furniture make the CFR vary over time. In Fig.~\ref{fig:pearsonr} we report two examples of CFR collected in the empty bedroom on two different days and the Pearson correlation coefficient [15] computed for each pair of traces acquired in the empty bedroom on different days and averaged over the monitoring antennas. The Pearson coefficient takes values in the range [-1, 1] where zero means that the traces are uncorrelated, while an absolute value of 1 means that the traces are linearly dependent. Both the qualitative plots and the quantitative evaluation show that the CFR traces are almost uncorrelated, confirming the need for domain robustness.

The unprecedented domain diversity offered by our dataset enables (i) the analysis of the statistics of the Wi-Fi CFR under many different domains; and (ii) the design and evaluation of the effectiveness of machine and deep learning algorithms for wireless sensing activity recognition, person identification and people counting that leverage domain-independent features, like, e.g., SHARP~\cite{SHARP-PAPER}, or domain-adaptive techniques, like, e.g., one-shot or few-shot learning~\cite{Restuccia}.
Specifically, we identified seven different use cases as detailed below.\\
\noindent\textbf{WHS robustness over time.} CFR measurements collected on many different days in the same environment and involving the same subject are included in the dataset, e.g, \texttt{AR-1a-e}, \texttt{AR-5a-b}. The analysis of this data allows understanding how small environmental changes impact the Wi-Fi channel statistics, enabling the design of algorithms that are robust to such changes.\\
\noindent\textbf{WHS robustness over environments and people.} Channel recordings have been collected by having the same subject performing activities in different environments, e.g., \texttt{AR-1a-e}, \texttt{AR-5a-b} and \texttt{AR-6a}, and having different subjects performing activities in the same (\texttt{AR-1a} and \texttt{AR-2a}) or different environments (\texttt{AR-7a} and \texttt{AR-8a}). This notable diversity is very useful to test AR algorithms that are to be integrated into commercial devices, to offer plug-and-play sensing functionalities and environment/person adaptability.\\
\noindent\textbf{WHS robustness over Wi-Fi devices.} Our dataset allows evaluating the performance of algorithms trained on CFR collected when data traffic is generated by Netgear routers on CFR collected when Asus routers are transmitting and vice versa. This allows studying methodologies to enforce transmission hardware robustness. Preliminary results on our SHARP algorithm~[4] indicate that these domain changes (hardware) deserve attention and further research.\\
\noindent\textbf{WHS domain adaptation.} The high domain diversity in the dataset allows designing domain adaptive algorithms. This is usually accomplished through learning-based approaches and entails setting initial model parameters based on a specific environment and refining them in case domain changes occur. The refinement is performed by retraining some parameters based on a few examples of data from the new domain, as proposed in, e.g.,~\cite{Restuccia}.\\
\noindent\textbf{Impact of multi-path and interference.} The measurements collected in the semi-anechoic chamber allow gauging how the same person modifies the multi-path from a controlled and almost multi-path-free semi-anechoic chamber (\texttt{AR-9a-c}) to a much more multi-path affected office environment (\texttt{AR-8a-b}). Moreover, the dataset has been mostly collected in environments where other \mbox{Wi-Fi} devices were operating beyond our control, thus causing radio interference. The impact of such interference can be evaluated by comparing the traces collected in the semi-anechoic chamber (no interference) with those collected in the other environments.\\
\noindent\textbf{Impact of obstructions.} Our dataset allows evaluating the robustness of WHS techniques when the path between the transmitter and the monitor is obstructed. This can be done for AR (\texttt{AR-3a-b}, \texttt{AR-4a}), PI (\texttt{PI-2a}, \texttt{PI-4a}) and PC (\texttt{PC-2a}, \texttt{PC-4a}) applications.\\
\noindent\textbf{Impact of \texttt{Tx} and \texttt{M} router locations.} The dataset includes traces {\it simultaneously} collected by two different devices (\texttt{M3} and \texttt{M4}) that concurrently monitor two transmission links (sets \texttt{PI}, \texttt{PC}). This allows evaluating how the position of the \texttt{Tx}-\texttt{M} link affects the performance of WHS algorithms.\\
\noindent\textbf{Impact of time, frequency, and spatial diversity.} New research avenues include analyzing the impact of the sampling time on the sensing accuracy by sub-sampling the CFR traces and using them as input for sensing algorithms. A second aspect regards analyzing whether sensing can benefit from selecting a sub-set of the collected OFDM sub-channels that may be more relevant for specific tasks (see e.g.,~\cite{meneghello2023toward}). Lastly, the impact of the number of monitoring antennas can be assessed (see e.g.,~\cite{SHARP-PAPER}).

The provided examples are only some of the possible ways to use the dataset. We trust that the variety of available settings will spur new reproducible research.

\section{Conclusions}
\label{conclusion}

In this paper, we have presented a new comprehensive dataset that allows developing robust and domain-adaptive learning-based models for human sensing, making it possible to train and test the algorithms over different conditions in terms of deployed \mbox{Wi-Fi} devices, person/people involved, and/or environment. This allows tackling a common pitfall of existing techniques, being that they {\it do not generalize well} as the online working conditions change with respect to those considered in their training phase. Our dataset provides a common ground for performance assessment, allowing for reproducible research. 
Future work will include pairing Wi-Fi CFR records obtained through equally spaced antenna arrays and multiple devices with webcam data that provides the ground truth for the subjects' location, thus providing support for developing localization and tracking applications.

\section*{Acknowledgment}
This work was partially supported by the European Union under the Italian National Recovery and Resilience Plan (NRRP) of NextGenerationEU, partnership on “Telecommunications of the Future” (PE0000001 - program “RESTART”) and by the European Union's Horizon 2020 programme under Grants No. 871249, project LOCUS. The views/opinions are those of the authors and do not necessarily reflect those of the funding institutions.

\bibliographystyle{IEEEtran}
\bibliography{IEEEabrv,biblio}

\vspace{-1cm}
\begin{IEEEbiographynophoto} 
{Francesca Meneghello} is an Assistant Professor at the University of Padova, Italy. Her research interests include \mbox{deep-learning} architectures and signal processing with application to wireless communications and sensing.
\end{IEEEbiographynophoto}
\vspace{-1cm}

\begin{IEEEbiographynophoto}
{Nicol\`o Dal Fabbro} is a Ph.D. student at the University of Padova, Italy. His research interests are in optimization over networks and wireless sensing.
\end{IEEEbiographynophoto}
\vspace{-1cm}

\begin{IEEEbiographynophoto} 
{Domenico Garlisi} is an Assistant Professor at the University of Palermo, Italy. He has been a Visiting Researcher at the Department of Computer Science, UCLA. His main research is related to wireless networks, software-defined radio, sensor networks, IoT and wireless systems for user localization and testing.
\end{IEEEbiographynophoto}
\vspace{-1cm}

\begin{IEEEbiographynophoto} 
{Ilenia Tinnirello} is a Full Professor at the University of Palermo, Italy. She has been Visiting Researcher at the Seoul National University, Korea, and the Nanyang Technological University of Singapore. Her research activities focus on wireless networks, and in particular on the design and prototyping of protocols and architectures for emerging reconfigurable wireless networks. She has been involved in several European research projects.
\end{IEEEbiographynophoto}
\vspace{-1cm}

\begin{IEEEbiographynophoto}
{Michele Rossi} is a Full Professor at the University of Padova, Italy. His research interests include wireless sensing, green mobile networks, edge and wearable computing and the use of machine and deep learning algorithms for signal processing. He currently coordinates the EU ITN project GREENEDGE (GA no. 953775) on green edge computing. He currently serves on the Editorial Boards of the IEEE Transactions on Mobile Computing, and of the IEEE Open Journal of the Communications Society.
\end{IEEEbiographynophoto}
\vfill

\end{document}